\pdfoutput=1

\documentclass[floatfix,showpacs,amsmath,amssymb,pra,twocolumn]{revtex4}

\usepackage{dsfont}
\usepackage{rotating}
\usepackage{subfigure}

 \input{Qcircuit}

\newcommand{\ketbra}[2]{\ensuremath{\ket{#1}\bra{#2}}}

\newcommand{\HS}[1]{\ensuremath{\mathcal{#1}}} 

\newcommand{\tr}{\ensuremath{\mathrm{tr}}}
\newcommand{\Tr}{\tr}
\newcommand{\id}{\mathds{1}}

\newcommand{\Real}{\ensuremath{\mathds{R}}}
\newcommand{\Cplx}{\ensuremath{\mathds{C}}}

\newcommand{\halmos}{\newline\vspace{3mm}\hfill $\Box$}
\newcommand{\proof}{\noindent {\it Proof.\ }}

\newtheorem{lemma}{Lemma}

\newcommand{\ie}{\emph{ie.}}
\newcommand{\etal}{\emph{et al.}}
\newcommand{\eJ}{\emph{J}}
\newcommand{\eS}{\emph{S}}

\usepackage{xcolor}

\begin{document}

\title{Experimentally feasible measures of distance between quantum operations}
\date{3 November 2009} 

\author{Zbigniew Pucha{\l}a}
\email{z.puchala@iitis.pl}

\author{Jaros{\l}aw Adam Miszczak}
\email{miszczak@iitis.pl}

\author{Piotr Gawron}
\email{gawron@iitis.pl}

\author{Bart{\l}omiej Gardas}
\email{bgardas@iitis.pl}
\affiliation{Institute of Theoretical and Applied Informatics, Polish Academy
of Sciences, Ba{\l}tycka 5, 44-100 Gliwice, Poland}

\begin{abstract}
We present two measures of distance between quantum processes based on the
superfidelity, introduced recently to provide an upper bound for quantum
fidelity. We show that the introduced measures partially fulfill the
requirements for distance measure between quantum processes. We also argue that
they can be especially useful as diagnostic measures to get preliminary
knowledge about imperfections in an experimental setup. In particular we provide
quantum circuit which can be used to measure the superfidelity between quantum
processes.

As the behavior of the superfidelity between quantum processes is crucial for
the properties of the introduced measures, we study its behavior for several
families of quantum channels. We calculate superfidelity between arbitrary
one-qubit channels using affine parametrization and superfidelity between
generalized Pauli channels in arbitrary dimensions. Statistical behavior of the
proposed quantities for the ensembles of quantum operations in low dimensions
indicates that the proposed measures can be indeed used to distinguish quantum
processes.
\end{abstract}
\maketitle


\section{Introduction}
Recent applications of quantum mechanics are based on processing and
transferring information encoded in quantum states~\cite{BZ06,hayashi}. The full
description of quantum information processing procedures is given in terms of
\emph{quantum channels} or \emph{quantum processes}, \ie\ completely positive,
trace non-increasing maps on the set of quantum states~\cite{BZ06}. 

In many areas of quantum information processing one needs to quantify the
difference between ideal quantum procedure and the procedure which is performed
in the laboratory. This is especially true in the situation when one deals with
imperfections during the realization of experiments. 
Theoretically these imperfections can be measured using state
tomography~\cite{leonhardt96discrete,jones91principles} or process tomography
\cite{poyatos97complete,ariano01tomography}. In particular the problem of
quantifying the distance between quantum channels was studied in the context of
channel distinguishability \cite{acin01statistical, ariano05minimax,
wang06unambiguous, watrous08distinguishing}.

The problem of identifying a universal measure which could be used for this
purpose was first comprehensively addressed in~\cite{gilchrist05distance}. In
this work the authors provided the list of requirements which should be
satisfied theoretically, as well as experimentally, in order to make the
measures of distance between quantum processes meaningful. They identified four
quantities which could be used as such measure, namely fidelity (\emph{J
fidelity} or $F_{\mathrm{pro}}$) and trace distance (\emph{J~process distance}
or $D_{\mathrm{pro}}$) between Jamio{\l}kowski matrices representing processes,
stabilized process fidelity (\emph{S~fidelity} or $F_{\mathrm{stab}}$) and
stabilized process distance (\emph{S~distance} or
$D_{\mathrm{stab}}$)~\cite[Sec.~IV.C]{gilchrist05distance}.

Both \eS~fidelity and \eS~distance are based on the optimization procedures with
respect to the set of quantum states. They share the common drawback: it is hard
to provide a~general formula for calculating any of those quantities. Formulas 
where given only in some special cases \cite{aharonov97mixed,Ji2006}. Numerical
calculation of those quantities can be reduced to a~convex optimization problem,
but it still requires time consuming process tomography. The main advantage of
those quantities is their appealing physical interpretation. On the other hand,
such measures are also hard to handle from the mathematical point of view. 

Also \eJ~fidelity and \~process distance are relatively hard to measure in a
laboratory as both can be calculated only after the full process tomography. In
the case of \eJ~fidelity simpler procedure can be given only in the case when
one aims to compare a unitray evolution with an arbitrary process. Hovever, even
in this case one needs to estimate $\mathcal{O}(d^2)$ observable everages for
$d$-dimensional quantum system~\cite{gilchrist05distance}.

The main aim of this paper is to present two measures of distance between quantum
processes based on the \emph{superfidelity}, introduced recently to provide an
upper bound for quantum fidelity \cite{subsuper,mendonca,PM09}, to the problem
of quantifying distance between quantum channels. We introduce metrics on the
space of quantum operations based on superfidelity and we examine their
properties. We also propose a simple quantum circuit which allows for the
measurement of superfidelity between quantum processes. Hence, to our knowledge,
we provide the first examples of metrics on the space of quantum operations
which can be measured directly in laboratory without resorting to process
tomography. We test our quantities against the requirements introduced
in~\cite{gilchrist05distance} and show their relations with \eJ~fidelity
introduced therein. We argue that the proposed metrics can be especially useful
as the diagnostic measures allowing to get preliminary knowledge about
imperfections in an experimental setup. 

This paper is organized as follows. In Section~\ref{sec:qintro} we recall basic
facts concerning distance measures on the space of quantum states and methods of
generalization to the space of quantum processes. In Section
\ref{sec:super-process} we show that distinguishability measures on the set of
quantum channels constructed using superfidelity partially fulfill the
requirements stated in \cite{gilchrist05distance}. Moreover, we propose, as
a~direct generalization of results obtained in~\cite{subsuper}, the quantum
circuit for measuring superfidelity between quantum processes. In Section
\ref{sec:examples} we provide exemplary analysis of several quantum channels
using the introduced measures and in Section~\ref{sec:stat-analysis} we discuss
statistical properties of introduced quantities. Finally, in Section
\ref{sec:final} we summarize the presented work and provide some concluding
remarks.

\section{Quantum states and operations}\label{sec:qintro}
Let $\HS{H}$ be a~separable, complex Hilbert space used to describe the system
in question. In quantum information theory we deal mainly with
finite-dimensional Hilbert spaces, so usually we are in the situation where
$\HS{H}=\Cplx^n$. The state of the system is described by the density matrix, \ie\
operator $\rho:\HS{H}\rightarrow\HS{H}$, which is positive ($\rho\geq0$) and
normalized ($\tr{\rho}=1$). 

In what follows we denote by $\mathcal{M}_N$ the space of density matrices of
size~$N$. We restrict our attention to the finite-dimensional case.

\subsection{Distance measures between quantum states}
In many situations in quantum information theory it is important to quantify to
what degree states are similar to the average state or how, on average, the
given quantity evolves during the execution of quantum procedure. The crucial
question emerging in this situation is how one should choose random states from
the set of density matrices. This is equivalent to choosing how one should
measure distance between quantum states. As a density matrix is the analogue of
the classical probability distribution, one can find among distance measures
quantifying distance between quantum state analogues of classical
quantities~\cite{BZ06}.

Among the mostly used metrics we can point out the trace distance, the
Hilbert-Schmidt distance, and the Bures distance. Bures distance is the most
natural one used in the analysis of quantum states. It has many important
properties~\cite{BZ06}. In particular it is a Riemannian and monotone metric. On
the space of pure states it reduces to Fubini-Study metric~\cite{Uh76} and it
induces statistical distance in the subspace of diagonal density matrices.

For the sake of consistency we introduce the following convention to denote
distance measures on $\mathcal{M}_N$. Let $X$ be a~functional on
$\mathcal{M}_N$. We denote by $A_X$, $B_X$ and $C_X$ the following quantities
\begin{eqnarray}
A_X &=& \arccos \sqrt{X}, \\
B_X &=& \sqrt{2-2\sqrt{X}}, \\
C_X &=& \sqrt{1-X},
\end{eqnarray}
which are motivated by \emph{Bures angle}, \emph{Bures distance} \cite{BZ06} and
\emph{root infidelity} \cite{gilchrist05distance}.

For $\rho,\sigma\in\mathcal{M}_N$ Bures distance can be defined in terms of
quantum fidelity \cite{Uh76} as
\begin{equation}\label{eqn:def-bures}
B_F(\rho,\sigma) = \sqrt{2-2\sqrt{F(\rho,\sigma)}}.
\end{equation}
Here $F$ is quantum fidelity 
\begin{equation}
F(\rho,\sigma)= \left[\tr|\sqrt{\rho}\sqrt{\sigma}|\right]^2,
\end{equation}
which provides the measure of similarity on the space of density matrices.

For two density matrices $\rho_1, \rho_2\in\mathcal{M}_N$ the trace distance is
defined as
\begin{equation}
D_{\tr}(\rho_1,\rho_2)=\frac{1}{2}\tr|\rho_1-\rho_2|.
\end{equation}

Recently a new measure of similarity between quantum states, namely
\emph{superfidelity} $G(\rho,\sigma)$, was introduced~\cite{subsuper}
\begin{equation}\label{eqn:def-superfidelity}
G(\rho,\sigma)=\tr\rho\sigma + \sqrt{1-\tr\rho^2}\sqrt{1-\tr\sigma^2}.
\end{equation}

The most interesting feature of the superfidelity is that it provides an upper
bound for quantum fidelity \cite{subsuper}
\begin{equation}\label{eqn:F<G}
F(\rho_1,\rho_2)\leq G(\rho_1,\rho_2),
\end{equation}
and a bound for the trace distance \cite{PM09}
\begin{equation}
1-D_{\tr}(\rho_1,\rho_2) \leq G(\rho_1,\rho_2).
\end{equation}
In (\ref{eqn:F<G}) we have an equality either for $\rho,\sigma\in\mathcal{M}_2$
or in the case where one of the states is pure. 

The superfidelity has also properties which make it useful for quantifying the
distance between quantum states. In particular we have:
\begin{enumerate}
\item Bounds: $0 \le G(\rho_1,\rho_2) \le 1$.
\item Symmetry: $G(\rho_1,\rho_2)= G(\rho_2,\rho_1)$.
\item Unitary invariance: for any unitary operator $U$, we have
  $G(\rho_1,\rho_2)=G(U\rho_1U^{\dagger},U\rho_2U^{\dagger})$.
\item Concavity: $G(\rho_1,\alpha \rho_2 + (1-\alpha)\rho_3) \geq \alpha
  G(\rho_1,\rho_2) + (1-\alpha) G(\rho_1,\rho_3)$ for any
  $\rho_1,\rho_2,\rho_3\in\Omega_N$ and $\alpha \in [0,1]$.
\item Supermultiplicativity: for $\rho_1,\rho_2,\rho_3,\rho_4 \in \Omega_N$ we
  have
  \begin{equation}
  G(\rho_1 \otimes \rho_2, \rho_3 \otimes \rho_4) \geq G(\rho_1,\rho_3)
  G(\rho_2,\rho_4).
  \end{equation}
\end{enumerate}

Note that the superfidelity shares properties 1.-4. with fidelity. However, in
contrast to the fidelity, the superfidelity is not multiplicative, but
supermultiplicative.

In \cite{mendonca} the authors showed that $G$ is \emph{jointly} concave in its
two arguments. Note that the property of joint concavity is obeyed by square
root of the fidelity but not by the fidelity.

It was also shown that it can be used to define such metrics on $\mathcal{M}_N$
\cite{subsuper} as 
\begin{equation}\label{eqn:dist-CG}
 C_G(\rho,\sigma) = \sqrt{1 - G(\rho,\sigma)}
\end{equation}
or
\begin{equation}\label{eqn:dist-arccos}
 A_{G^2}(\rho_1,\rho_2) = \arccos(G(\rho_1,\rho_2)).
\end{equation}

Before we discuss further properties of these metrics in the context of quantum
channels we should note that the function defined as
\begin{equation}
 B_G(\rho,\sigma) = \sqrt{2 - 2\sqrt{G(\rho,\sigma)}}
\end{equation}
is not a~metric \cite{mendonca}, and thus it is impossible to provide a direct
generalization of Bures distance in terms of superfidelity. Also, in contrast to
the fidelity or trace distance, $G$ is not monotone \cite{mendonca}, thus
neither $C_G$ nor $A_{G^2}$ can be studied using Morozova-\v{C}encov-Petz
theorem \cite[Ch.~14]{BZ06}.

\subsection{Quantum processes}
The most general form of the evolution of a quantum system is given in terms of
quantum channels. In this paper we consider quantum channels which are 
Completely Positive Trace Preserving (CP-TP) maps. 

In order an map $\Phi$ to be a CP-TP map it has to fulfill set of the following
conditions:
\begin{enumerate}
\item It has to preserve trace, positivity and hermiticity, \ie\
  \begin{equation}
	  \Tr{\Phi(\rho)}=1, \Phi(\rho)\geq0\ \mathrm{and\ } \Phi(\rho)=\Phi(\rho)^\dagger.
  \end{equation}
\item It has to be linear
  \begin{equation}
	  \Phi\left(\sum_i p_i\rho_i \right)=\sum_i p_i \Phi\left(\rho_i \right).
  \end{equation}
\item Finally it has to be \emph{completely positive}, \ie\ for $\rho^{(n)} \geq
  0$ we require that
  \begin{eqnarray}
	   (\Phi\otimes\id_n) \rho^{(n)} \geq 0,\ n\in N,
  \end{eqnarray}
  where $\rho^{(n)}$ is an element of an appropriate space of 
  states.
\end{enumerate}
These conditions are required for $\Phi$ to preserve the set of quantum states.

In the most general case quantum evolution is described by a superoperator
$\Phi$, acting on $\mathcal{M}_{N}$, which can be expressed in Kraus form
\cite{BZ06, hayashi}
\begin{equation}
\Phi(\rho)=\sum_k E_k^{\phantom{\dagger}} \rho E_k^\dagger,
\end{equation}
where $\sum_k E_k^\dagger E_k^{\phantom{\dagger}}=\id$.

Alternatively quantum operations an be represented by a \emph{superoperator
matrix} $M_\Phi$. The superoperator matrix is a~representation of linear
operator in the canonical basis. The following formula allows to transform set
of Kraus operators $\{E_k\}$ into superoperator matrix $M_\Phi$~\cite[Ch.
10]{BZ06}
\begin{equation} \label{r:Mphi}
	M_\Phi=\sum_{k=1}^{N^2} E_k \otimes E^\ast_k, 
\end{equation}
where $N = \dim(E_k)$ and `$\ast$' denotes element-wise complex conjugation.

The dynamical matrix for the operations $\Phi$ is defined as $D_\Phi=M_\Phi^R$,
where `${}^R$' denotes a \emph{reshuffling} operation~\cite{BZ06}. The dynamical matrix
for the trace preserving operation acting on $N$-dimensional system is
an~$N^2\times N^2$ positive defined matrix with trace~$N$. We can introduce
natural correspondence between such matrices and density matrices on $N^2$ by
normalizing $D_\Phi$. Such a correspondence is known as \emph{Jamio{\l}kowski
isomorphism}~\cite{jamiolkowski72linear, zyczkowski04duality}.

Let $\Phi$ be a~completely positive trace preserving map acting on density
matrices. We define Jamio\l{}kowski matrix of $\Phi$ as
\begin{equation} \label{r:jam}
 \rho_\Phi = \frac{1}{N}D_\Phi. 
\end{equation}

Jamio\l{}kowski matrix has the same mathematical properties as a quantum state
\ie{} it is a semi-definite positive matrix with trace equal to one. It is
sometimes referred to as \emph{Jamio\l{}kowski state matrix}.

\subsection{Distance measures between quantum processes}\label{sec:super-process}
The problem of finding the measure of difference between ideal and real quantum
processes was first studied in depth in~\cite{gilchrist05distance}, where the
authors proposed the list of requirements for gold-standard metric between
quantum processes.

If $\Delta$ is a candidate for distance measure, the criteria are as follows:
\newcounter{requirement}
\begin{list}{(R\arabic{requirement})}{\usecounter{requirement}}
 \item \emph{Metric}: $\Delta$ should be a metric.\label{req:metric}
 \item \emph{Easy to calculate}: it should be possible to evaluate $\Delta$
 in a direct manner.\label{req:easycalculate}
 \item \emph{Easy to measure}: there should be a clear and achievable
 experimental procedure for determining the value of
 $\Delta$.\label{req:easymeasure}
 \item \emph{Physical interpretation}: $\Delta$ should have a
 well-motivated physical interpretation.\label{req:physicalinterpret}
 \item \emph{Stability}: $\Delta(\id \otimes \Phi, \id \otimes \Psi) =
\Delta(\Phi, \Psi)$, where $\id$ is the identity operation on an additional
quantum system.\label{req:stability} 
 \item \emph{Chaining}: $\Delta(\Phi_2 \circ \Phi_1,\Psi_2 \circ \Psi_1) \leq
 \Delta(\Phi_1 ,\Psi_1) + \Delta(\Phi_2 ,\Psi_2)$.\label{req:chaining}
\end{list}

As already noted in~\cite{gilchrist05distance}, it is hard to find a quantity
which fulfills all of the above requirements. On contrary, in many cases it is
desirable to use some kind of quantity which does not posses all of the required
features to get some preliminary insight into the nature of errors occurring in
the experimental setup.

\section{Distance measures based on superfidelity}
Let $\Delta_G$ be a distance measure based on the superfidelity between
Jamio{\l}kowski states of processes. In this paper we consider two functions
$C_G$, motivated by root infidelity, 
\begin{equation}
C_G(\Phi,\Psi) = \sqrt{1-G(\rho_\Phi,\rho_\Psi)},
\end{equation}
and $A_{G^2}$, motivated by Bures angle,
\begin{equation}
A_{G^2}(\Phi,\Psi) = \arccos G(\rho_\Phi,\rho_\Psi).
\end{equation}
We argue that both quantities seem to be suitable for metrics on the space of
processes.

\subsection{Basic properties (R\ref{req:metric}, R{\ref{req:easycalculate}})}
It was shown in~\cite{subsuper} that quantities defined in
Eqs.~(\ref{eqn:dist-CG}) and~(\ref{eqn:dist-arccos}) do provide the~metrics
on the space of quantum states. As such $C_G$ and $A_{G^2}$ fulfill requirement
(R\ref{req:metric}).

Also from the definition of superfidelity it is clear that $\Delta_G$ can be
easily calculated --- requirement (R\ref{req:easycalculate}). From the
computational point of view the calculation of $\Delta_G$, using standard
mathematical software, is also much efficient than in the case of metrics based
on fidelity~\cite{mendonca}.

\subsection{Measurement procedure (R\ref{req:easymeasure}) and 
physical~interpretation~(R\ref{req:physicalinterpret})}
Any useful distance measure for quantum processes should be easy to measure in
a laboratory. In the case of any metric based on superfidelity this is to say
that it should be easy to measure the superfidelity between quantum processes.

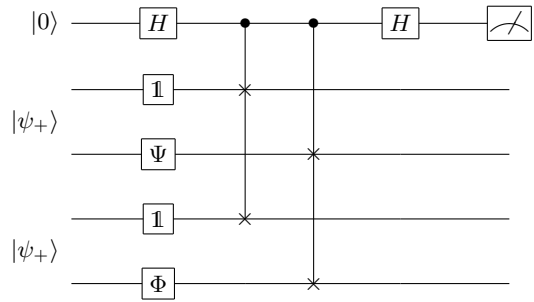
\begin{figure}[htp!]
\[
\Qcircuit @C=2.8em @R0.7em {
\lstick{\ket{0}}      & \gate{H}    & \ctrl{6} & \ctrl{8} & \gate{H}   & \meter\\
                      &             &                     &            & \\
                      & \gate{\id}  & \qswap   & \qw      &\qw         & \qw\\
\lstick{\ket{\psi_+}} &             &                     &            & \\
                      & \gate{\Psi} & \qw      &  \qswap  &\qw         & \qw\\
                      &             &                     &            & \\
                      & \gate{\id}  & \qswap   & \qw      &\qw         & \qw\\
\lstick{\ket{\psi_+}} &             &                     &            & \\
                      & \gate{\Phi} & \qw      & \qswap   &\qw         & \qw
}
\]
\caption{Quantum circuit for measuring $\tr \rho_\Phi\rho_\Psi$. The probability
$P_0$ of finding the top qubit in state $\ket{0}$ leads to an estimation of $\tr
\rho_\Phi\rho_\Psi = 2 P_0 - 1$ \cite{ekert02direct}. This allows direct
estimation of process superfidelity.}
\label{fig:measure-circ}
\end{figure}

In~Fig.~\ref{fig:measure-circ} a quantum circuit used for measuring the
superfidelity between quantum processes is presented. In the first step one
needs to produce Jamio{\l}kowski matrices for analyzed processes as described
in~\cite{ariano01tomography}. In the second step we utilize the scheme proposed
in~\cite{ekert02direct}.

The circuit works for quantum channels of an arbitrary dimension. Its only
drawback is that it requires controlled SWAP operation, which makes it
problematic for realization using contemporary technology
\cite{banaszek-private}.

In order to measure the superfidelity between two one-qubit channels one needs
$5$ qubits and for measuring the superfidelity between two $n$-dimensional
states one needs $2+4n$ dimensional space.

One should also note that the presented quantum circuit can be used to measure
the fidelity between a unitary operation and an arbitrary quantum channel. As
such it can be used in the situation when one needs to measure the difference
between an ideal (\ie\ unitary) process and a real (\ie\ noisy) process. In this
case we can easily give physical interpretation of the superfidelity between
quantum processes \cite{gilchrist05distance}, while in general case the
interpretation is still not clear.

\subsection{Stability (R\ref{req:stability})}
In this paragraph we show the \emph{stability} of distance measures based on
superfidelity $\Delta_G$ between Jamio{\l}kowski matrices of processes. In fact
we will even show, that if we extend both channels by the same unitary channel
(not necessarily identity) the superfidelity-based distance measures do not
change. 

We have the following lemma.
\begin{lemma}\label{lem:unitary-extension}
Let $\Psi,\Phi$ be given channels and let $\tau$ be a unitary quantum channel,
then
\begin{equation}
 G(\rho_{\tau \otimes \Psi}, \rho_{\tau \otimes \Phi} )
 = G(\rho_{\Psi}, \rho_{\Phi} ).
\end{equation}
\end{lemma}
\proof
To prove the above all we need is the fact that Jamio{\l}kowski state of unitary
channel is a rank 1 projector, the fact that $\rho_{\Psi_1 \otimes \Psi_2}$ is a
permutation similar to $\rho_{\Psi_1} \otimes \rho_{\Psi_2}$ and the following
lemma.
\halmos

\begin{lemma}
 Let $\ket{\phi}$ be a normalized vector, then
\begin{equation}
 G(\ketbra{\phi}{\phi} \otimes \rho_1, \ketbra{\phi}{\phi} \otimes \rho_2)
= G(\rho_1,\rho_2).
\end{equation}
\end{lemma}
\proof
To obtain the lemma it is enough to notice that 
\begin{eqnarray}
&& \tr (\ketbra{\phi}{\phi} \otimes \rho_i)( \ketbra{\phi}{\phi} \otimes \rho_j)\\\nonumber
&& = \tr (\ketbra{\phi}{\phi} \ketbra{\phi}{\phi}) \tr \rho_i \rho_j = \tr \rho_i \rho_j
\end{eqnarray}
for any $i,j \in \{1,2\}$.
\halmos

From Lemma~\ref{lem:unitary-extension} we have that any $\Delta_G$ fulfills
requirement~(R5).

\subsection{Chaining (R\ref{req:chaining})}
Despite its simple form superfidelity, in contrast to fidelity or trace
distance, is not monotone under the action of quantum channels. This fact was
proved in~\cite{mendonca}. One can easily construct an example similar to the
one used in~\cite{mendonca} to see that the superfidelity between quantum
channels fails to fulfill requirement (R\ref{req:chaining}).

Let us consider the following Jamio\l{}kowski states
\begin{equation}
\rho_{\Phi_1} = \rho_{\Phi_2} = 
\frac{1}{2}
\left(
\begin{array}{cccc}
 1 & 0 & 0 & 0 \\
 0 & 1 & 0 & 0 \\
 0 & 0 & 0 & 0 \\
 0 & 0 & 0 & 0
\end{array}
\right), 
\end{equation}
\begin{equation}
\rho_{\Psi_1} =
\frac{1}{2}
\left(
\begin{array}{cccc}
 1 & 0 & 0 & 0 \\
 0 & 0 & 0 & 0 \\
 0 & 0 & 0 & 0 \\
 0 & 0 & 0 & 1
\end{array}
\right), \
\rho_{\Psi_2} =
\frac{1}{2}
\left(
\begin{array}{cccc}
 0 & 0 & 0 & 0 \\
 0 & 0 & 0 & 0 \\
 0 & 0 & 1 & 0 \\
 0 & 0 & 0 & 1
\end{array}
\right)
\end{equation}
representing quantum channels $\Phi_1,\Phi_2,\Psi_1$ and $\Psi_2$ respectively.
Jamio\l{}kowski matrices corresponding to the compositions of
$\Phi_1\circ\Psi_1$ and $\Phi_2\circ\Psi_2$ read~\cite{BZ06}
\begin{equation}
\rho_{\Psi_1 \circ \Phi_1} =
\frac{1}{2} \left(
\begin{array}{cccc}
 1 & 0 & 0 & 0 \\
 0 & 1 & 0 & 0 \\
 0 & 0 & 0 & 0 \\
 0 & 0 & 0 & 0
\end{array}
\right)
\end{equation}
and
\begin{equation}
\rho_{\Psi_2 \circ \Phi_2}=
\frac{1}{2}
\left(
\begin{array}{cccc}
 0 & 0 & 0 & 0 \\
 0 & 0 & 0 & 0 \\
 0 & 0 & 1 & 0 \\
 0 & 0 & 0 & 1
\end{array}
\right).
\end{equation}
The superfidelity between the above Jamio\l{}kowski matrices reads
\begin{equation}
 G(\rho_{\Phi_1},\rho_{\Phi_2})=1, \ G(\rho_{\Psi_1} ,\rho_{\Psi_2} )= 3/4
\end{equation}
and
\begin{equation}
 G(\rho_{\Psi_1 \circ \Phi_1},\rho_{\Psi_2 \circ \Phi_2})=1/2.
\end{equation}
Taking this into account we get
\begin{eqnarray}
\frac{1}{\sqrt{2}} = 
C_G(\rho_{\Psi_1 \circ \Phi_1},\rho_{\Psi_2 \circ \Phi_2})\nleq \\\nonumber
C_G(\rho_{\Phi_1},\rho_{\Phi_2})+ C_G(\!\!\!\!&\rho_{\Psi_1},\rho_{\Psi_2}) = 
\frac{1}{2}.
\end{eqnarray}
We also have
\begin{eqnarray}
\frac{\pi}{3} = 
A_{G^2}(\rho_{\Psi_1 \circ \Phi_1},\rho_{\Psi_2 \circ \Phi_2})\nleq \\\nonumber
A_{G^2}(\rho_{\Phi_1},\rho_{\Phi_2})+ A_{G^2}(\!\!\!\!&\rho_{\Psi_1},\rho_{\Psi_2}) = 
\arccos{\frac{3}{4}}.
\end{eqnarray}
Thus the chaining rule does not hold for metrics $G_G$ and $A_{G^2}$ in the
general case.	

However, this property holds if we aim to compare unitary (\ie\ ideal) quantum
operations with general (\ie\ noisy) quantum operations. In this particular case
superfidelity reduces to \eJ~fidelity.

Chaining rule is important if one aims to compare quantum processes divided into
smaller steps. It holds for distance measures proposed
in~\cite{gilchrist05distance}.

\section{Examples}\label{sec:examples}
To get a deeper insight into a behavior of superfidelity-based distances 
we provide explicit formulas for the selected families of quantum channels.

\subsection{One-qubit channels}
We start by analyzing one-qubit channels. In this case dynamical matrix can
parametrized as~\cite{fujiwara99parametrization} (up to two orthogonal
transformations \cite[Sec.~10.7]{BZ06})
\begin{equation}
D=\frac{1}{2}\left(
\begin{smallmatrix}
 \eta_z+\kappa_z+1 & 0 & \kappa_x+i \kappa_y & \eta_x+\eta_y \\
 0 & -\eta_z+\kappa_z+1 & \eta_x-\eta_y & \kappa_x+i \kappa_y \\
 \kappa_x-i \kappa_y & \eta_x-\eta_y & -\eta_z-\kappa_z+1 & 0 \\
 \eta_x+\eta_y & \kappa_x-i \kappa_y & 0 & \eta_z-\kappa_z+1
\end{smallmatrix}
\right),
\end{equation}
where parameters $\vec{\kappa}=(\kappa_x,\kappa_y,\kappa_z)$ and
$\vec{\eta}=(\eta_x,\eta_y,\eta_z)$ are real vectors representing distortion and
translation of the quantum state in the Bloch ball.

Let $D_\Psi$ and $D_\Phi$ be two dynamical matrices parametrized by vectors
$\vec{\kappa}_\Psi$, $\vec{\eta}_\Psi$ and $\vec{\kappa}_\Phi$,
$\vec{\eta}_\Phi$ respectively.

After straightforward calculations we get 
\begin{align}
&G(\rho_\Psi, \rho_\Phi)=\frac{1}{4}\big(1+\vec{\kappa}_\Psi\cdot\vec{\kappa}_\Phi+\vec{\eta}_\Psi\cdot\vec{\eta}_\Phi +\\
\nonumber
&\sqrt{3-||\vec{\kappa}_\Psi||^2-||\vec{\eta}_\Psi||^2}\sqrt{3-||\vec{\kappa}_\Phi||^2-||\vec{\eta}_\Phi||^2}
\big),
\end{align}
where `$\cdot$' denotes the scalar product.

One should note that it is hard to obtain concise formula for the fidelity or
trace distance between two one-qubit channels.

One of the simplest examples of one-qubit maps are unital maps, \ie\ quantum
operations that transform maximally mixed state into itself. One-qubit unital
channels are exactly those with $\vec{\kappa}=(0,0,0)$. In this case we can
derive the formula for fidelity
\begin{align}
&F(\rho_\Psi,\rho_\Phi)=\\
\nonumber
&\frac{1}{16} \Big(\sqrt{(\eta_x^\Psi-\eta_y^\Psi-\eta_z^\Psi+1) (\eta_x^\Phi-\eta_y^\Phi-\eta_z^\Phi+1)}\\
\nonumber
&+\sqrt{(\eta_x^\Psi+\eta_y^\Psi-\eta_z^\Psi-1) (\eta_x^\Phi+\eta_y^\Phi-\eta_z^\Phi-1)}\\
\nonumber
&+\sqrt{(\eta_x^\Psi-\eta_y^\Psi+\eta_z^\Psi-1) (\eta_x^\Phi-\eta_y^\Phi+\eta_z^\Phi-1)}\\
\nonumber
&+\sqrt{(\eta_x^\Psi+\eta_y^\Psi+\eta_z^\Psi+1) (\eta_x^\Phi+\eta_y^\Phi+\eta_z^\Phi+1)}\Big)^2,
\end{align}
and for trace distance
\begin{align}
&D_{\tr}(\rho_\Psi,\rho_\Phi)=\\
\nonumber
&\frac{1}{8}\Big(|\eta_x^\Psi+\eta_y^\Psi+\eta_z^\Psi-\eta_x^\Phi-\eta_y^\Phi-\eta_z^\Phi|\\
\nonumber
&+|\eta_x^\Psi-\eta_y^\Psi+\eta_z^\Psi-\eta_x^\Phi+\eta_y^\Phi-\eta_z^\Phi|\\
\nonumber
&+|\eta_x^\Psi+\eta_y^\Psi-\eta_z^\Psi-\eta_x^\Phi-\eta_y^\Phi+\eta_z^\Phi|\\
\nonumber
&+|\eta_x^\Psi-\eta_y^\Psi-\eta_z^\Psi-\eta_x^\Phi+\eta_y^\Phi+\eta_z^\Phi|\Big).
\end{align}
between $\Psi$ and $\Phi$.

\subsection{Selected higher-dimensional channels}
We start with an elementary result concerning the superfidelity on commuting
matrices \cite{subsuper}.
\begin{lemma}
Let $\rho_1$ and $\rho_2 $ be hermitian matrices with eigenvalues
$\vec{\lambda}$ and $\vec{\mu}$ respectively. If $\rho_1 \rho_2 = \rho_2 \rho_1$
then there exists an orthonormal basis $\{\ket{i}\}_i$ such that 
\begin{equation}
 \rho_1 = \sum_i \lambda_i \ketbra{i}{i}\quad\mathrm{and}\quad
 \rho_2 = \sum_i \mu_i \ketbra{i}{i}.
\end{equation}
With this notation we have  
\begin{equation}
	G(\rho_1,\rho_2) = 
	\vec{\lambda} \cdot \vec{\mu} + \sqrt{(1-|\vec{\lambda}|^2)(1-|\vec{\mu}|^2)}.
\end{equation}
\end{lemma}

This lemma enables us to obtain explicit formulas for the superfidelity between
quantum channels for some interesting families discussed below.

\subsubsection{Depolarizing channel}
For any $p\in [0,1]$ we define a depolarizing channel as~\cite{hayashi}
\begin{equation}
 \kappa_{d,p}(\rho) = p \rho + (1-p) \tr (\rho) \frac{1}{d} \id .
\end{equation}
It is a $d$-dimensional CP-TP map. It is not difficult to notice that
$\rho_{\kappa_{d,p}}$ and $\rho_{\kappa_{d,q}}$ commute, and eigenvalues of
$\rho_{\kappa_{d,p}}$ are
\begin{equation}
 \left \{\underbrace{\frac{1-p}{d^2},\frac{1-p}{d^2},\dots, 
 \frac{1-p}{d^2}}_{d^2-1}, \frac{1}{d^2}(1+(d^2-1)p) \right\}.
\end{equation}
Thus we have 
\begin{align}
&G(\rho_{\kappa_{d,p}},\rho_{\kappa_{d,q}})= \\
\nonumber
&\frac{1}{d^2} \left( 1 + (d^2-1) p q + (d^2-1)\sqrt{(1-p^2)(1-q^2)}  \right).
\end{align}

\subsubsection{Generalized Pauli channel}
Generalized Pauli channel $\Pi_d$ is an extension to any dimension of the
one-qubit Pauli channel \cite{hayashi}. We define two families of unitary
operators:
\begin{equation}
	X_d=\sum_0^{d-1} \ket{j-1\bmod d}\bra{j},
\end{equation} 
and 
\begin{equation}
	Z_d=\mathrm{diag}\left(1,e^{2i \pi/d \times 1},\ldots e^{2i \pi/d \times (d-1)}\right).
\end{equation} 
The channel action is defined as
\begin{equation}
\Pi_d(\rho)=\sum_{i,j=0}^{d-1} p_{i,j} X_d^i Z_d^j \rho (X_d^i Z_d^j)^\dagger,
\end{equation}
where $0\leq p_{i,j}\leq 1$ and $\sum p_{i,j}=1$.

For two generalized Pauli channels $\rho_p$ and $\rho_q$ given by the
probability distribution matrices $p_{i,j}$ and $q_{i,j}$, we can find a~direct
formula for their similarity in terms of superfidelity
\begin{equation}
	G(\rho_p,\rho_q)=\tr(pq^T)+\sqrt{1-\tr(pp^T)}\sqrt{1-\tr(qq^T)}.	
\end{equation}
This follows from the fact that $\rho_p$ and $\rho_q$ commute and vectors
$\vec{p},\vec{q} \in \Real^{d^2}$ are eigenvalues of $\rho_p$ and $\rho_q$
respectively.

\subsubsection{Werner-Holevo channel}
Werner-Holevo channel cannot be represented as generalized Pauli channel.

For dimension $d$ and parameter $p\in [-\frac{1}{d-1},\frac{1}{d+1}]$ we define
Werner-Holevo channel as 
\begin{equation}
 \kappa^{T}_{d,p}(\rho) = p \rho^{T} + (1-p) \tr (\rho) \frac{1}{d} \id .
\end{equation}
It is a $d$-dimensional CP-TP map and it is sometimes called transpose
depolarizing channel. Also in this case it is not difficult to notice that
$\rho^{T}_{\kappa_{d,p}}$ and $\rho^{T}_{\kappa_{d,q}}$ commute, and
eigenvalues of $\rho^{T}_{\kappa_{d,p}}$ are
\begin{eqnarray}
\Big\{
 \underbrace{\frac{1-(d+1)p}{d^2},\dots, \frac{1-(d+1)p}{d^2}}_{\binom{d}{2}},\\
 \underbrace{\frac{1+(d-1)p}{d^2},\dots, \frac{1+(d-1)p}{d^2}}_{\binom{d+1}{2}}
\Big\}.
\end{eqnarray}
Thus we have 
\begin{align}
&G(\rho^{T}_{\kappa_{d,p}},\rho^{T}_{\kappa_{d,q}})= \\
\nonumber
&
\frac{1}{d^2} \left( 1 + (d^2-1) p q + (d^2-1)\sqrt{(1-p^2)(1-q^2)}  \right).
\end{align}

Since the dynamical matrices for depolarizing channel and Werner-Holevo channel
commute, one can also easily calculate the superfidelity between these channels.
In this case it reads
\begin{align}
&G(\rho_{\kappa_{d,p}},\rho^{T}_{\kappa_{d,q}})= \\
\nonumber
&
\frac{1}{d^2} \left( 1 + (d-1) p q + (d^2-1)\sqrt{(1-p^2)(1-q^2)}  \right).
\end{align}

\subsubsection{Dephasing channel}\label{subsec:dephasing}
The most general dephasing~\cite{Alicki} \ie\ pure decoherence channel for a
single qubit is defined as
\begin{equation}\label{r:50}
D_{f_{{t}}} : \rho_0\rightarrow\rho_t, 
\end{equation}
where
\begin{equation}\label{r:51}
\rho_t =
\left(
\begin{array}{cc}
 \rho_{{11}} & \rho_{{12}}f_{{t}}  \\
 \rho_{{12}}^{\ast}f_{{t}}^{\ast} & \rho_{{22}} 
\end{array}
\right),
\quad t\ge 0.
\end{equation}

For a given function $f_t$ channel $D_{f_t}$ is CP-TP quantum operation. This is
not obvious, but it follows from the fact that $D_{f_{{t}}}$ can be written in
Kraus representation as
\begin{equation}
D_{f_{t}}(\rho_0)=E_1(t)^{\dagger}\rho_0 E_1(t)+E_2(t)^{\dagger}\rho_0 E_2(t), \label{r:dc}
\end{equation}
where time-dependent Kraus operators $E_i(t)$, $(i=1,2)$ are given by
\begin{equation}
 E_1(t)=\mbox{diag}(1,f_{t}),  \label{kr:1}
\end{equation}
and
\begin{equation}
E_2(t)=\mbox{diag}(0,\sqrt{1-|f_{t}|^2}). \label{kr:2}
\end{equation}

The function $f_t:\Real\rightarrow\Cplx$ was chosen in the way that $f_0=1$ and
$|f_t|\leq 1$. The last condition guarantees that the channel $D_{f_t}$ is a CP
quantum operation. An explicit formula for $f_t$ depends on the particular
choice of an environment which is used to model process of decoherence
(see~\cite{dephasing, asymmetric}).

Let $D_{f_{t}}$ and $D_{g_{t}}$ be given dephasing channels. According
to~(\ref{kr:1}) and (\ref{kr:2}), it has the following Kraus representation:  
\begin{equation}
 X_1(t)=\mbox{diag}(1,x_t),\ X_2(t)=\mbox{diag}(0,\sqrt{1-|x_t|^2}), \label{r:kraus}
\end{equation}
where $x=f,g$. Let us denote the Jamio{\l}kowski matrix associated with channel
$D_{x_t}$ by $\rho_{x_t}$. Then we can easily see that
\begin{equation}
 \rho_{x_{{t}}}=\frac{1}{2}
 \left(
\begin{array}{cccc}
 1 & 0 & 0 & x_t  \\
 0 & 0 & 0 & 0  \\
 0 & 0 & 0 & 0 \\
 x_t^{\ast} & 0 & 0 & 1 
\end{array}
\right),
\quad x=f,g,
\end{equation}
therefore the superfidelity $G_t(f,g):=G(\rho_{f_{t}},\rho_{g_{t}})$
reads
\begin{equation}
G_t(f,g)=
\frac{1}{2}+\frac{1}{2}\Re(f_tg_t^{\ast})+\frac{1}{2}\sqrt{1-|f_t|^2}\sqrt{1-|g_t|^2}.
\label{r:sf}
\end{equation}
Here $\Re$ stands for the real part of a complex number. An interesting
situation arises when $g_t=f_t^{\ast}$. In this case from~(\ref{r:sf}) we have
\begin{equation}
 G_t(f,f^{\ast})=1-\frac{|f_t^2|-\Re(f_t^2)}{2}.
\end{equation}

The above considerations can be easily generalized for the arbitrary
qu\emph{d}it (\ie\ $d$-dimensional state). Indeed, let $F_t=F_t^{\dagger}$ be a
$d$-dimensional \emph{dephasing matrix} \ie\ $(F_t)_{ii}=1$ and
$(F_t)_{ij}=f_{ij}(t)$ for $i\not=j$.
We define channel $D_{F_t}$ as follows 
\begin{equation}
 D_{F_t}:\rho_0\rightarrow F_t \bullet \rho_0, \label{r:bullet} 
\end{equation}
 where by `$\bullet$' we denoted the Hadamard product of matrices. 
One can easily see that for these types of channels
\begin{equation}
 G_t(\vec{f},\vec{g})=\frac{1}{d^2}\left((\vec{f}_t)^{\dagger}\cdot\vec{g}_t
    +\sqrt{d^2-\|\vec{f}_t\|^2}\,\sqrt{d^2-\|\vec{g}_t\|^2},\right)
\label{r:gen1}
\end{equation}
which in the case $\vec{g}_t=\vec{f}_t^{\ast}$ reduces to 
\begin{equation}
G_t(\vec{f},\vec{f}^{\ast}) = 1 - 
\frac{\|f_t\|^2-(\vec{f}_t)^{\dagger}\cdot \vec{f}_t^{\ast}}{d^2}. \label{r:gen2} 
\end{equation}
Here $\vec{f}_t$ stands for a vector obtained from matrix $F_t$ by the reshaping
procedure~\cite{BZ06}, and $\|\cdot\|$ represents a standard norm on
$\mathbb{C}^d$.

Note also that the results~(\ref{r:gen1}) and~(\ref{r:gen2}) hold in the case
of arbitrary hermitian matrix $F_t$, not only a dephasing matrix.

\section{Statistical analysis of channels}\label{sec:stat-analysis}
In order to asses the quality of the distance measures based on superfidelity we
have analyzed its statistical behavior. We have compared the average
superfidelity with the average fidelity between one-qubit quantum channels. We
have also analyzed average superfidelity and average fidelity between quantum
channels for higher-dimensional random channels.

\subsection{Benchmarks for one-qubit channels}
Measures based on fidelity and trace distance (\eJ~fidelity and \eJ~process
distance) provide natural benchmarks for testing new measures on the space of
quantum operations.

\begin{figure}[t!]
\begin{center}
\subfigure[Probability distributions of fidelity and superfidelity.]{
  \includegraphics[width=\columnwidth]{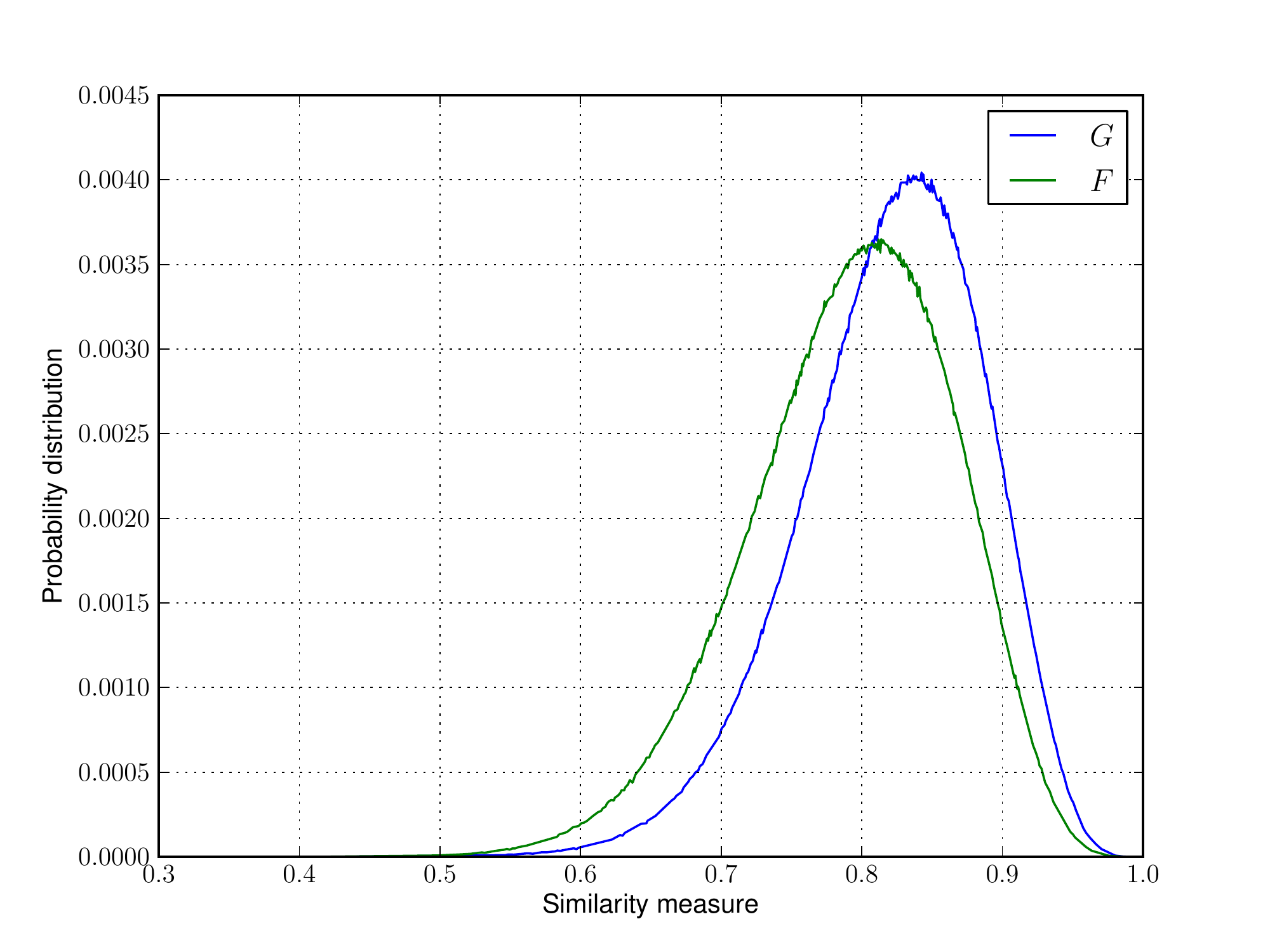}
  \label{fig:fidelity-super-qubit}  
}
\subfigure[Probability distributions of trace distance ($D_\tr$), root
infidelity ($C_F$) and root ``superinfidelity'' ($C_G=\sqrt{1-G}$).]{
  \includegraphics[width=\columnwidth]{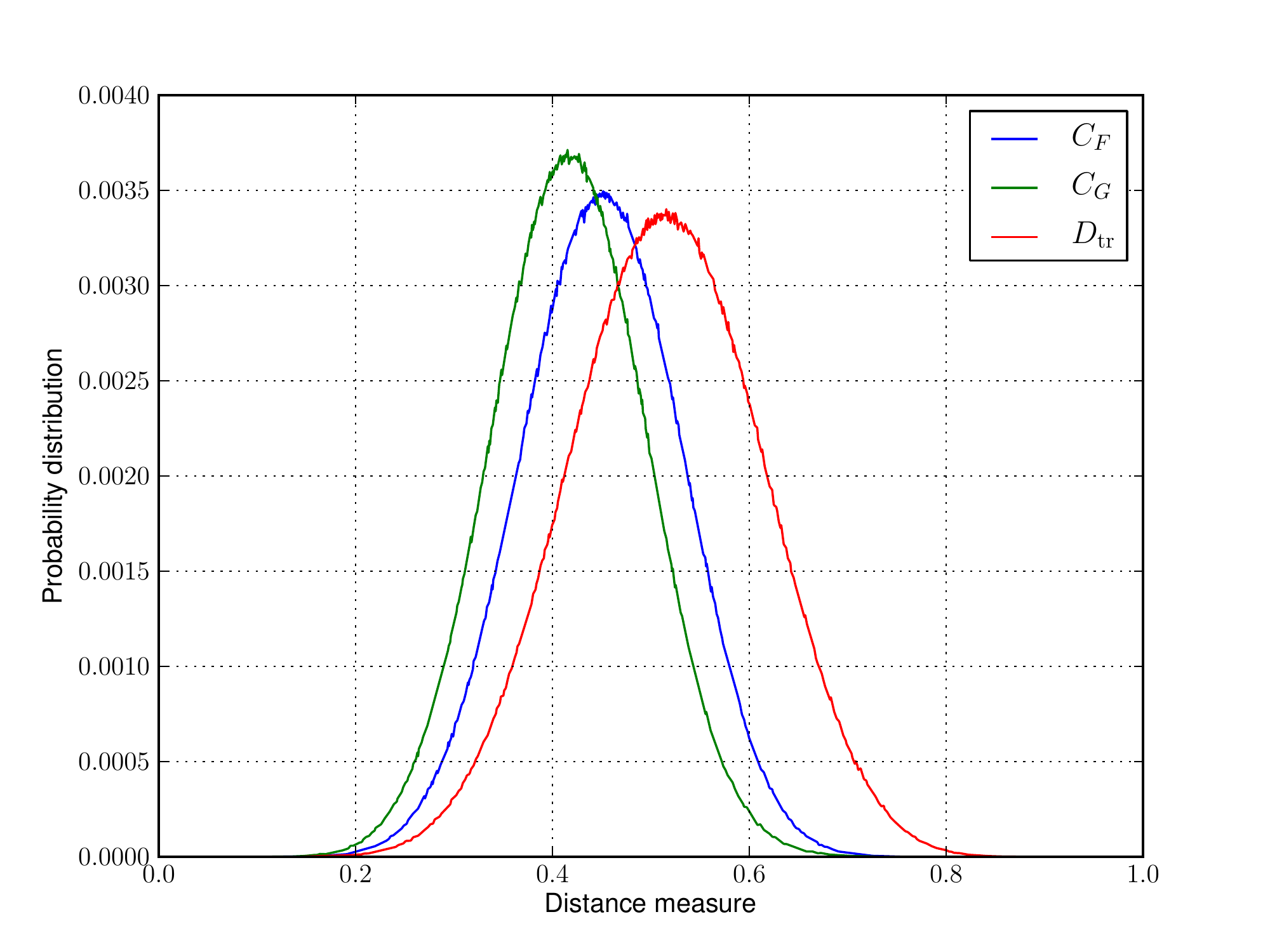}
  \label{fig:measures-compare-qubit}
}
\caption{Statistical comparison of (a) fidelity and superfidelity and (b)
distance measures for one-qubit quantum channels.}
\label{fig:benchmarks-qubit}
\end{center}
\end{figure}

Using the algorithm by Bruzda \etal\ \cite{bruzda00operations} we have generated
$10^6$ pairs of normalized dynamical matrices representing one-qubit quantum
channels. For this sample we have calculated the fidelity and the superfidelity (see
Fig.~\ref{fig:fidelity-super-qubit}) and distance measures $C_F$, $C_g$,
$D_\mathrm{tr}$ (see Fig.~\ref{fig:measures-compare-qubit}).

Numerical results presented in Fig.~\ref{fig:measures-compare-qubit} indicate
that in the case of one-qubit channels the superfidelity (or metrics based on
it) can be used to approximate trace distance or measures based on fidelity.
Thus, the circuit used to measure the superfidelity can be used to provide some
insight into the behavior of these measures.

\subsection{Average superfidelity between channels}
\begin{figure}[t!]
\begin{center}
\subfigure[Fidelity]{
	\includegraphics[width=\columnwidth]{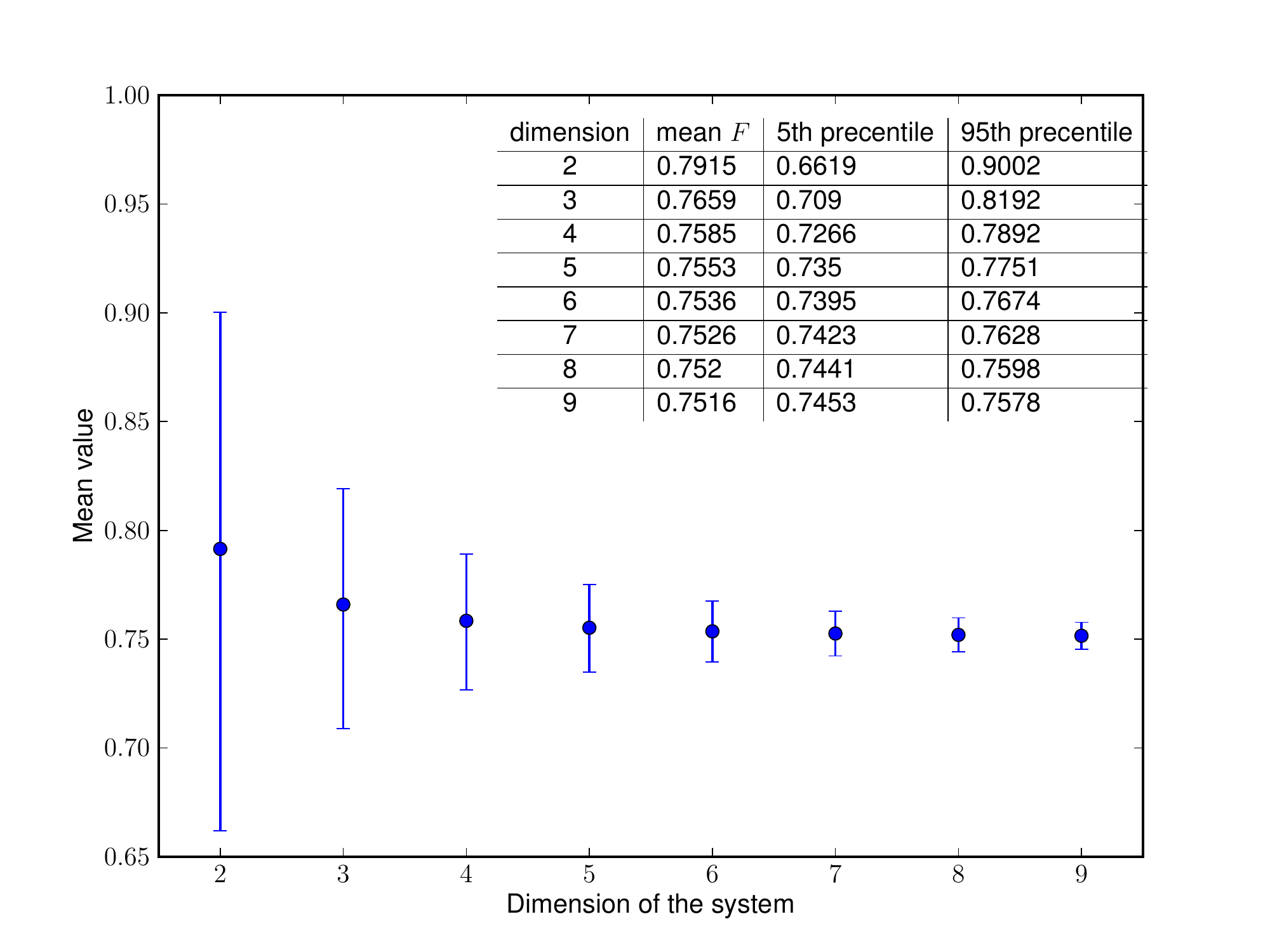}
	\label{fig:mean-fidelity}
}
\subfigure[Superfidelity]{
	\includegraphics[width=\columnwidth]{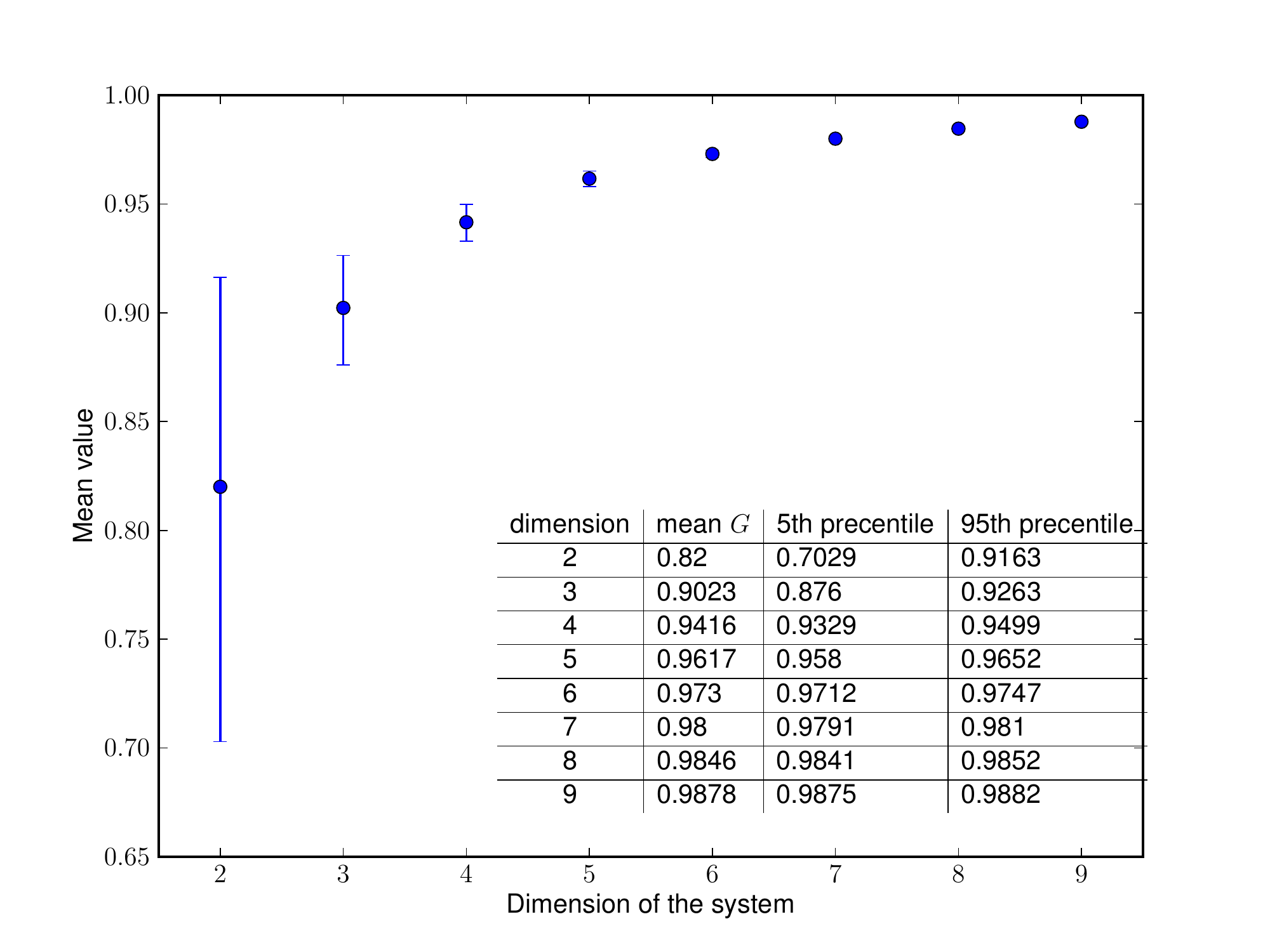}
	\label{fig:mean-superfidelity}
}
\caption{Mean, 5th percentile and 95th percentile of similarities between random
quantum operations acting on the state space of dimensions 2 to
9.}\label{fig:mean}
\end{center}
\end{figure}

In order to describe which maps are distant it is helpful to seek the average
behavior of the superfidelity between quantum maps. Having this information we
can judge which maps are distant by comparing the superfidelity between them
with the quantils over the space of quantum operations. Fig.~\ref{fig:mean}
shows mean fidelity and mean superfidelity between quantum channels together
with 5th and 95th percentile for the channels that act on qudits of dimensions
two to nine.

\section{Concluding remarks}\label{sec:final}
We have introduced the measure of similarity between quantum processes
constructed as the superfidelity between corresponding Jamio{\l}kowski states.
We have also used this quantity to introduce two metrics on the space of quantum
operations -- $C_G$ and $A_G$ -- motivated by root infidelity and Bures angle.
We have argued that the introduced quantities can be used as diagnostic measures
for probing errors occurring during physical realizations of quantum information
processing. This is especially true as we have shown that the presented
quantities can be potentially measured in laboratory. Also, a quantum circuit,
constructed to measure the superfidelity. can be used to measure the fidelity
between a unitary evolution, regarded as an ideal channel, and an arbitrary
quantum process, realized in a laboratory. Thus, the presented quantum circuit
can be used to calibrate experimental setup with respect to some ideal setup.
For the special case of one-qubit channels superfidelity between quantum
operations can be used as a~relatively good approximation of the fidelity.

\textbf{Acknowledgements} We acknowledge the financial support by the Polish
Ministry of Science and Higher Education under the grant number N519 012
31/1957 and Polish Research Network LFPPI.

We wish to thank Wojtek Bruzda for his Matlab code for generating random
dynamical matrices and Karol \.Zyczkowski, Wojtek Roga and Mario Ziman for
helpful remarks and inspiring discussions.

Numerical calculations presented in this work were performed on
the \texttt{Leming} server of The Institute of Theoretical and Applied
Informatics, Polish Academy of Sciences.


\end{document}